\documentclass[12pt]{article}
\usepackage{amsmath}
\usepackage{cite}
\usepackage{epsfig}
\usepackage{epstopdf}
\usepackage{amssymb}
\usepackage{bbold}

\usepackage{setspace}

\def\[{\left [}
\def\]{\right ]}
\def\({\left (}
\def\){\right )}

\def\r{\rho}

\def\r2{\sqrt{2}}

\newcommand{\bbibitem}[1]{\bibitem{#1}\marginpar{#1}}

\def\Label#1{\label{#1}%
  \smash{\hbox to0pt{\raise1ex\hbox{\tiny[#1]}\hss}}}
\def\noLabels{\let\Label=\label}
\def\nobbibitem{\let\bbibitem=\bibitem}

\newcommand{\bea}{\begin{eqnarray}}
\newcommand{\eea}{\end{eqnarray}}
\newcommand{\beq} {\begin{equation}}
\newcommand{\eeq} {\end{equation}}
\newcommand{\beqa} {\begin{eqnarray}}
\newcommand{\eeqa} {\end{eqnarray}}

\newcommand{\beqn}{\begin{eqnarray}}
\newcommand{\eeqn}{\end{eqnarray}}

\begin{document}

\begin{flushright}
OUTP-14-04P
\end{flushright}

\vskip 2cm \centerline{\Large {\bf Chern-Simons interactions in AdS$_3$ and}}
\vskip .5cm \centerline{\Large {\bf the current conformal block }}
 \vskip 1.5cm
\renewcommand{\thefootnote}{\fnsymbol{footnote}}
\centerline
{{\bf Ville Ker\"anen,
\footnote{Ville.Keranen@physics.ox.ac.uk}
}}
\vskip .5cm
\centerline{\it Rudolf Peierls Centre for Theoretical Physics,}
\centerline{\it University of Oxford, 1 Keble Road,}
\centerline{\it Oxford OX1 3NP, United Kingdom}

\setcounter{footnote}{0}
\renewcommand{\thefootnote}{\arabic{footnote}}

\begin{abstract}
We compute the four point function of scalar fields in AdS$_3$ charged under $U(1)$ Chern-Simons fields
using the bulk version of the operator state mapping. Then we show how this four point function is 
reproduced from a CFT$_2$ with a global $U(1)$ symmetry, through the contribution of the corresponding current
operator in the operator product 
expansion, i.e. through the conformal block of the current operator. We work in a "probe approximation"
where the gravitational interactions are ignored, which corresponds to leaving out the energy
momentum tensor from the operator product expansion.

\end{abstract}

\onehalfspacing

\section{Introduction}

Chern-Simons (CS) theory provides a beatiful example of a holographic duality. Chern-Simons theory on a
manifold with a boundary can be directly related to a 1+1 dimensional CFT on the boundary, 
the Wess-Zumino-Witten (WZW) model \cite{Witten:1988hf,Elitzur:1989nr}. The gauge symmetry of the bulk Chern-Simons theory
is related to the global symmetry of the WZW model, much like in the AdS/CFT duality.

In the context of the AdS$_3$/CFT$_2$ duality, Chern-Simons gauge fields naturally
arise in the bulk as the low energy limit of string theory  \cite{Gukov:2004ym,Gukov:2004id}. 
As is well known, even gravity in AdS$_3$ (at least classically) can be written as
a Chern-Simons theory \cite{Achucarro:1987vz,Witten:1988hc}. As discussed recently in 
\cite{Jensen:2010em}, much of  the structure of the original CS/WZW duality 
generalize to the examples arising in the AdS/CFT duality, although important
differences also appear \cite{Gukov:2004id}.

One particular new feature that arises in the AdS/CFT examples is that generically one
has dynamical matter in the bulk charged under the Chern-Simons fields. 
The purpose of this note is to study the correlation functions of CFT operators
dual to scalar fields in the bulk charged under the CS fields. A nice feature of the 
Chern-Simons interaction is that the calculation
of the three and four point functions is considerably simpler (one might even say trivial),
than in higher dimensional examples (see e.g. \cite{D'Hoker:1999pj}). 
Thus, they provide simple examples of higher point correlation functions
that are easily calculable in AdS.

The original CS/WZW duality is particularly nice as the "dictionary" between
the bulk and the boundary is explicit. Thus, bulk results can be 
straightforwardly understood from the boundary perspective as well.
This is unlike in general examples of AdS/CFT where even simple bulk 
physics, such as locality for example, can be complicated from the
boundary theory perspective (even though considerable recent progress has been made
towards understanding bulk locality see e.g. \cite{Heemskerk:2009pn,Fitzpatrick:2012yx}).
We find that this simplicity at least partially
continues to the case with dynamical charged matter, as we can give an
interpretation of the Chern-Simons interaction in the four point function 
very simply within the CFT. The CS interaction can be
reproduced using the operator product expansion (OPE) as follows. The
CFT has a current operator dual to the CS gauge fields. The contribution of
the current operator and its conformal descendants to the OPE can be resummed
to give rise to what is called the conformal block of the current operator. 
The current conformal block precisely reproduces the effect of the bulk CS
interaction from the CFT. This is unlike in more general examples of the 
AdS/CFT duality \cite{ElShowk:2011ag}, where the conformal block of a given
operator is known not to be enough to reproduce the exchange interaction of the
corresponding dual bulk field.

Throughout this note we will work in a "probe approximation" where
bulk gravity is ignored, and also we treat the Chern-Simons interaction perturbatively
in powers of $1/k$, where $k$ is the Chern-Simons level. Thus, we are assuming $c\gg k \gg 1$,
where $c$ is the central charge of the CFT.
Also we only consider $U(1)$ Chern-Simons theory, while generalization to non-Abelian groups
seems straightforward.

The current manuscript is organized as follows.
In section 2 we provide a short introduction to classical
Chern-Simons theory in AdS$_3$, and compute two point functions of the corresponding dual current operators. 
In section 3 we discuss the operator state mapping in the context of AdS/CFT, which will be used
in the rest of the paper. In section 4 we compute a three point function of charged scalars and the current
as a warm up. In section 5 we compute the four point function of charged scalar operators. 
In section 6, we introduce the OPE formalism in the CFT and show that the current operator conformal
block reproduces the bulk four point function. In section 7 we perform a simple check of the results
by computing the interaction energy between charged bulk particles and compare the
result to the CFT anomalous dimension. In section 8 we discuss the relation of the CS interaction
to more complicated bulk interactions.

\section{Review of Chern-Simons Holography}

The Euclidean action of a single Chern-Simons field is
\beq
S_E=-\frac{k}{8\pi}\int d^3x\epsilon^{\mu\nu\lambda}A_{\mu}\partial_{\nu}A_{\lambda}+S_{bdy}.\label{eq:CS}
\eeq
Here we use complex coordinates $z=x+i\tau$, $\bar{z}=x-i\tau$ in terms of which the metric of $AdS_3$ is
\beq
ds^2=\frac{1}{u^2}(du^2+dzd\bar{z}).
\eeq
From this point on, we will use the gauge $A_u=0$. The non-vanishing gauge field components are thus $A_{z}$ and $A_{\bar{z}}$. The equations of motion $F_{\mu\nu}=0$, in this gauge imply
that $A_z$ and $A_{\bar{z}}$ are both independent of $u$ and that
\beq
F_{z\bar{z}}=0.\label{eq:CSeom}
\eeq
This can be solved by setting \footnote{Topologically Euclidean AdS$_3$ is the product of a disk and a real line. Thus,
all closed forms in AdS$_3$ are also exact.}
\beq
A_z=\partial_z\Lambda(z,\bar{z}),\quad A_{\bar{z}}=\partial_{\bar{z}}\Lambda(z,\bar{z}).\label{eq:sol}
\eeq
Now we have exhausted all the equations of motion. Next we should specify boundary conditions. Specifying both
$A_z$ and $A_{\bar{z}}$ as boundary conditions is overconstraining as for generic $A_z$ and $A_{\bar{z}}$ no $\Lambda(z,\bar{z})$
satisfying (\ref{eq:sol}) exists. Thus, we are lead to fix one combination of the two as a boundary condition.
We will choose to fix $A_{\bar{z}}$. In the following we will see that this is indeed a
consistent boundary condition with a simple physical meaning in the boundary CFT.

For the boundary condition to be consistent with the variational principle, we need to ensure that the boundary terms in the
variation of the action vanish. The variation of the Chern-Simons action is
\beq
\delta S_E=\frac{k}{8\pi}\int d^2z\Big( A_{\bar{z}}\delta A_z-A_z\delta A_{\bar{z}}\Big)+\delta S_{bdy}.
\eeq
Gauge field configurations within the space of allowed variations (respecting our boundary conditions) satisfy $\delta A_{\bar{z}}=0$, leading to the variation
\beq
\delta S_E=\frac{k}{8\pi}\int d^2z\delta A_z A_{\bar{z}}+\delta S_{bdy}.
\eeq
If we choose the following covariant boundary term
\beq
S_{bdy}=-\frac{k}{16\pi}\int d^2z\sqrt{|\gamma|}A^2=-\frac{k}{8\pi}\int d^2zA_z A_{\bar{z}},
\eeq
the variation of the action vanishes identically $\delta S_E=0$. If we had chosen the boundary condition
$\delta A_z=0$, the boundary term would have the opposite sign.

Within the holographic dictionary, $A_{\bar{z}}$ is identified as a background gauge field coupled to a left moving current $j_z$. According to the usual holographic dictionary
\beq
\langle j_z\rangle=\frac{\delta S}{\delta A_{\bar{z}}}=-\frac{k}{4\pi}A_z |_{u=0}.\label{eq:currentdictionary}
\eeq
The current two point function can be obtained from the current one point function in the presence of a delta function source for the gauge field. To see this
we write the current expectation value with the delta function source as
\beq
\langle j_z(z')\rangle_{A_{\bar{z}}=\delta^{(2)}(z)}=\frac{\langle j_z(z')e^{-j_z(0)}\rangle}{\langle e^{-j_z(0)}\rangle},
\eeq
where the latter expectation values are taken in the CFT ground state. Expanding the exponential and using large-$N$ factorization of the correlation
functions (which we are assuming here) gives
\beq
\langle j_z(z')\rangle_{A_{\bar{z}}=\delta^{(2)}(z)}=-\langle j_z(z')j_z(0)\rangle.\label{eq:1vs2}
\eeq
Thus, to obtain the holographic two point function, we need to find a solution to the Chern-Simons equations of motion that satisfies $A_{\bar{z}}=\delta^{(2)}(z)$. The corresponding solution is of the form (\ref{eq:sol}) with
\beq
\Lambda(z,\bar{z})=\frac{1}{2\pi z},\label{eq:soln}
\eeq
due to the identity $\partial_{\bar{z}}(1/(2\pi z))=\delta^{(2)}(z,\bar{z})$. The solution (\ref{eq:soln}) is not unique in the
sense that naively we could add an arbitrary function of $z$ to it, while satisfying the boundary conditions and the equation of motion.  
But requiring that the only singularity in $A_{\bar{z}}$ is a delta function at $z=0$, fixes the function of $z$ to be at most a constant.  
Thus, we obtain the current two point correlator
\beq
\langle j_z(z)j_z(0)\rangle=\frac{k}{4\pi}\partial_z\Lambda=-\frac{k}{8\pi^2}\frac{1}{z^2}.
\eeq
This is not quite the theory we want to consider in the following, since the current $j_{z}$ is not really conserved since (\ref{eq:currentdictionary}), 
together with the equation of motion (\ref{eq:CSeom}),  implies that
in the presence of a background gauge field one has
\beq
\langle \partial_{\bar{z}}j_z\rangle=-\frac{k}{4\pi}\partial_z A_{\bar{z}}.
\eeq
This is the chiral anomaly discussed in detail in \cite{Jensen:2010em}.

We would like to consider a case when the current in the dual field theory is fully conserved.
To achieve this, we should introduce another component $j_{\bar{z}}$ for the current to
obtain a parity invariant theory and to get rid of the anomaly. For this purpose we can introduce another gauge field
$\bar{A}$ interpreted as being dual to $j_{\bar{z}}$. The full action we choose is
\beq
S_E=-\frac{k}{8\pi}\int d^3x\epsilon^{\mu\nu\lambda}(A_{\mu}\partial_{\nu}A_{\lambda}-\bar{A}_{\mu}\partial_{\nu}\bar{A}_{\lambda})
-\frac{k}{8\pi}\int d^2z(A_z A_{\bar{z}}+\bar{A}_z \bar{A}_{\bar{z}}-2A_{\bar{z}}\bar{A}_{z}).\label{eq:CS2}
\eeq
For $\bar{A}$ to be dual to a right moving current, we fix $\bar{A}_z$ at the boundary. The $\bar{A}$ boundary term in (\ref{eq:CS2})
is chosen again to obtain a well defined variational principle. The coefficient of the last boundary term is chosen to
make the current conserved, as can be easily checked.

Again the current correlator is given by the current one point function
in the presence of a delta function source. The current correlation functions are now given by
\begin{align}
&G_{zz}=-\frac{k}{8\pi^2}\frac{1}{z^2},\quad G_{\bar{z}\bar{z}}=-\frac{k}{8\pi^2}\frac{1}{\bar{z}^2}.\nonumber
\\
&G_{z\bar{z}}=G_{\bar{z}z}=-\frac{k}{8\pi^2}\partial_z\Big(\frac{1}{\bar{z}}\Big).
\end{align}
Above we defined $G_{ab}=\langle j_a(z)j_b(0)\rangle$. The $G_{z\bar{z}}$ component appeared from the last boundary term
in (\ref{eq:CS2}).

Neglecting the contact term, the two point function can be written compactly in terms of cartesian coordinates on $R^2$ as
\beq
G_{ab}=\frac{C_V}{x^2}\Big(\delta_{ab}-2\frac{x_ax_b}{x^2}\Big),\quad C_V=\frac{k}{4\pi^2}.\label{eq:level}
\eeq

\section{Operator state mapping in AdS/CFT}

A CFT on the complex plane, with a complex coordinate $z$, can be mapped to the cylinder using the standard conformal map
\beq
w=\log z,\quad w=\tau+i\phi\label{eq:cylindermap1}
\eeq
Now a dilatation on the plane $z\rightarrow e^{\lambda} z$ takes the form of a time translation $\tau\rightarrow \tau+\lambda$ on the cylinder. 
Thus, the dilatation operator becomes the Hamiltonian on the cylinder.
Under the conformal transformation (\ref{eq:cylindermap1}), correlation functions of primary operators transform as
\beq
\langle\mathcal{O}_1(w_1,\bar{w}_1)...\mathcal{O}_n(w_n,\bar{w}_n)\rangle=\prod_{j=1}^{n}\Big(\frac{\partial z_j}{\partial w_j}\Big)^{h_j}\Big(\frac{\partial\bar{z}_j}{\partial \bar{w}_j}\Big)^{\bar{h}_j}\langle \mathcal{O}_1(z_1,\bar{z}_1)...\mathcal{O}_n(z_n,\bar{z}_n)\rangle,\label{eq:cylindermap2}
\eeq
where $(h_j,\bar{h}_j)$ are the conformal dimensions of the operators $\mathcal{O}_j$. The operators we will be mainly considering are
scalar operators for which $(h_j,\bar{h}_j)=(\Delta/2,\Delta/2)$, and the left moving current $j_z$ with $(h_j,\bar{h}_j)=(1,0)$ and the right moving
current $j_{\bar{z}}$ with $(h_j,\bar{h}_j)=(0,1)$. The CFT vacuum state corresponds to a Euclidean path integral with no operators located at $z=0$. Whenever an operator approaches $z=0$ it can be absorbed into
the definition of the initial state $\mathcal{O}(0)|0\rangle=|\mathcal{O}\rangle$. This is a dilatation eigenstate as can be seen from the conformal symmetry algebra, and thus, an energy eigenstate on the cylinder.
Similarly, as an operator approaches $z\rightarrow\infty$, it can be absorbed into the final state \cite{Belavin:1984vu,DiFrancesco:1997nk}
\beq
\lim_{z\rightarrow\infty}\bar{z}^{2h}z^{2\bar{h}}\langle 0|\mathcal{O}(\bar{z},z)=\langle \mathcal{O}|,\label{eq:finalstate}
\eeq
which is an energy eigenstate on the cylinder for the same reason.

In the context of AdS/CFT, a CFT on a plane is dual to AdS in the Poincare coordinates
\beq
ds^2=\frac{1}{u^2}(du^2+dzd\bar{z}),
\eeq
while a CFT on a cylinder is dual to AdS in global coordinates
\beq
ds^2=\frac{1}{\cos^2\theta}\Big(d\tau^2+d\theta^2+\sin^2\theta d\phi^2\Big).
\eeq
These two metrics are related by a coordinate transformation
\beq
\theta=\arctan\Big(\frac{|z|}{u}\Big),\quad w=\frac{1}{2}\log\Big[z^2\Big(1+\frac{u^2}{|z|^2}\Big)\Big],\label{eq:coordinatetransf}
\eeq
where $w=\tau+i\phi$. Indeed at the boundary $u\rightarrow 0$, the coordinate transformation reduces to the conformal
map (\ref{eq:cylindermap1}) from the plane to the cylinder.

To see how the operator state mapping works, consider a scalar operator inserted at $z=0$ in the CFT. In the leading order in the large-$N$ limit
the operator insertion is identical to turning on a delta function source for the corresponding operator at $z=0$.
In the bulk (Poincare patch), the solution approaching a delta function at th eboundary is the bulk to boundary propagator
\beq
\varphi_0(u,z,\bar{z})=C_{\Delta}\frac{u^{\Delta}}{(u^2+z\bar{z})^{\Delta}}.
\eeq
Inverting the coordinate transformation (\ref{eq:coordinatetransf})
\beq
|z|=\sin\theta e^{\tau},\quad u=\cos\theta e^{\tau},
\eeq
the bulk field configuration becomes\footnote{We have chosen the overall normalization so that $\varphi_0$ has
a unit Klein-Gordon norm.}
\beq
\varphi_0(\tau,\phi,\theta)=\frac{1}{\sqrt{2\pi}}e^{-\Delta\tau}(\cos\theta)^{\Delta}.\label{eq:wavefunction}
\eeq
When continued to real time, this is the wavefunction of the lowest energy eigenstate of the bulk scalar field in global AdS (see e.g. \cite{Fitzpatrick:2010zm}). When canonically quantizing the bulk scalar,
one uses the mode expansion
\beq
\varphi(x)=\sum_{n,l}(f_{n,l}(x)a_{n,l}+f_{n,l}^*(x)a_{n,l}^{\dagger}).
\eeq
The mode functions $f_{n,l}$ are labeled by two quantum numbers, the radial mode number $n$ and the angular momentum $l$.
The lowest energy state corresponds to the one we obtained through the operator state mapping $f_{0,0}=\varphi_0$.
We will denote the corresponding bulk quantum state as $a^{\dagger}_{0,0}|0\rangle=|1\rangle$.
The insertion of a primary operator corresponds to preparing the corresponding particle
in its lowest energy eigenstate. Thus, the holographic dictionary between the bulk and the boundary states is
\beq
|\mathcal{O}\rangle\quad \leftrightarrow \quad |1\rangle.
\eeq
The excited states created by $a^{\dagger}_{n,l}$ correspond to the (global) conformal descendant operators of the form
\beq
(\partial_{\mu}\partial^{\mu})^n\partial_{\mu_1}...\partial_{\mu_l}\mathcal{O}(0),\label{eq:descendantop}
\eeq
We refer the interested reader to \cite{Fitzpatrick:2010zm} for more details. Here we will only need the ground state wavefunction.
In 1+1 CFT in addition to (\ref{eq:descendantop}) there are also higher Virasoro descendants of the primary operators. In
the bulk, these correspond to adding boundary graviton excitations. We will work in the "probe approximation",
where these are ignored.

\section{A warm up three point function}

Next we couple a charged scalar field to the Chern-Simons theory. The action for the scalar field is chosen as
\beq
S_{scalar}=\int d^3x\sqrt{g}(D\varphi^\dagger D\varphi+m^2\varphi^\dagger\varphi),
\eeq
where $D_{\mu}\varphi=\partial_{\mu}\varphi-i \frac{e}{2}(A_{\mu}+\bar{A}_{\mu})\varphi$. In the following we will apply 
the operator state mapping to the three point function
\beq
G_3=\langle\mathcal{O}^{\dagger}(z_1,\bar{z_1})j_{z}(z_2,\bar{z_2})\mathcal{O}(z_3,\bar{z}_3)\rangle,
\eeq
where $\mathcal{O}$ is the operator dual to $\varphi$. The form of all three points functions in a CFT are fixed by conformal symmetry up to an overall constant. In the current case
\beq
G_3=\frac{C_{\mathcal{O}^{\dagger}j\mathcal{O}}}{2}\frac{1}{|z_{13}|^{2\Delta}}
\Big(\frac{1}{z_{23}}-\frac{1}{z_{21}}\Big).
\eeq
Above we introduced the notation $z_{ij}=z_i-z_j$.
Taking a limit where the scalar operators approach $z_3\rightarrow\infty$ and $z_1\rightarrow 0$ and using (\ref{eq:cylindermap2}) and (\ref{eq:finalstate}), we obtain the OPE coefficient in terms of a one point function on the cylinder
\beq
C_{\mathcal{O}^{\dagger}j\mathcal{O}}=2\langle\mathcal{O}|j_{w}|\mathcal{O}\rangle.\label{eq:j3point}
\eeq
Thus, we are lead to calculate the expectation value of the current operator $j_{w}$ in the lowest energy one particle state $|\mathcal{O}\rangle$.
This state in the bulk is simply
\beq
|1\rangle=a^{\dagger}_{0,0}|0\rangle,
\eeq
with the wavefunction (\ref{eq:wavefunction}). This state sources the gauge fields $A_{\mu}$ and $\bar{A}_{\mu}$ as dictated by the Chern-Simons equation of motion
\begin{align}
&\frac{k}{2\pi}\epsilon^{\mu\nu\lambda}\partial_{\nu}\bar{A}_{\lambda}=\sqrt{-g}J^{\mu},\label{eq:CSsourced}
\\
&\frac{k}{2\pi}\epsilon^{\mu\nu\lambda}\partial_{\nu}A_{\lambda}=-\sqrt{-g}J^{\mu},\label{eq:CSsourced2}
\end{align}
where $J^{\mu}$ is the current generated by the ground state wavefunction
\beq
J^{\mu}=-ig^{\mu\nu}(\varphi_0^*\partial_{\nu} \varphi_0-\varphi_0\partial_{\nu} \varphi_0^*),
\eeq
explicitly given by
\beq
J^{t}=\frac{e\Delta}{2\pi}(\cos\theta)^{2\Delta+2},\quad J^{\theta}=J^{\phi}=0.
\eeq
In (\ref{eq:CSsourced}) and (\ref{eq:CSsourced2}), we have neglected the $(A+\bar{A})^2\phi^{\dagger}\phi$ term as it gives contributions of the order $1/k^2$.
In the gauge $A_{\theta}=0$, the unique solution\footnote{Up to gauge transformations that vanish at the boundary and at the past and the future infinity.}
to equation (\ref{eq:CSsourced}), and the corresponding equation for $A_{\mu}$, respecting "normalizable" boundary conditions is
\begin{align}
&A=\frac{e}{k}dt+\frac{e}{k}(1-(\cos\theta)^{2\Delta})d\phi,\label{eq:A1}
\\
&\bar{A}=\frac{e}{k}dt-\frac{e}{k}(1-(\cos\theta)^{2\Delta})d\phi.\label{eq:A2}
\end{align}
In particular, as we approach the boundary, the gauge fields asymptote to
\beq
A=\frac{e}{k}(dt+d\phi),\quad \bar{A}=\frac{e}{k}(dt-d\phi).\label{eq:asygauge}
\eeq
Continuing to Euclidean time the boundary values of the fields become
\beq
A_w=-i\frac{e}{k}=\bar{A}_{\bar{w}}.
\eeq
Using the holographic dictionary we obtain
\beq
\langle\mathcal{O}|j_{w}|\mathcal{O}\rangle=-\frac{k}{4\pi}A_{w}=i\frac{e}{4\pi},\quad \langle\mathcal{O}|j_{\bar{w}}|\mathcal{O}\rangle=-\frac{k}{4\pi}\bar{A}_{\bar{w}}=i\frac{e}{4\pi}.
\eeq 
Comparison to (\ref{eq:j3point}) gives us the OPE coefficient
\beq
C_{\mathcal{O}^{\dagger}j\mathcal{O}}=i\frac{e}{2\pi},
\eeq
which is indeed the correct result as it is fixed
by a current conservation Ward identity as discussed in Appendix \ref{sec:C}. The bulk version of this statement
is that the asymptotic values of the gauge fields (\ref{eq:asygauge}) are fixed by the total charge in
the correspoding bulk state.

\section{Four point function of charged operators}\label{sec:4point}

The bulk version of the operator state mapping illustrated in the previous section becomes particularly
powerful in the case of 4-point functions. The operator state mapping has been earlier used in \cite{Hijano:2013fja} to
calculate four point functions in higher spin gravity in AdS$_3$. The following calculation is very similar in spirit
to that of \cite{Hijano:2013fja}. 

We will consider a system where in the bulk there are two scalar fields $\varphi_1$ and $\varphi_2$
that only interact through the Chern-Simons fields. Other interactions would give rise to additive corrections to 
our result, in the first order in bulk perturbation theory. We will consider the 4-point function of the corresponding dual scalar operators
\beq
G_4=\langle\mathcal{O}_1(z_1,\bar{z_1})\mathcal{O}_1^{\dagger}(z_2,\bar{z_2})\mathcal{O}_2^{\dagger}(z_3,\bar{z_3})\mathcal{O}_2(z_4,\bar{z_4})\rangle.\label{eq:4pointdef}
\eeq
This correlation function is neither fixed by conformal invariance nor by Ward identities. Conformal invariance guarantees that the
4-point function can be written in the (standard) form
\beq
G_4=\frac{1}{|z_{12}|^{2\Delta_1}|z_{23}|^{2\Delta_2}}g(u,v),\label{eq:F}
\eeq
where $z_{ij}=z_i-z_j$, and $u$ and $v$ are the (global) conformal invariant cross ratios
\beq
u=\Big|\frac{z_{12}z_{34}}{z_{13}z_{24}}\Big|^2,\quad v=\Big|\frac{z_{14}z_{23}}{z_{13}z_{24}}\Big|^2.
\eeq
Thus, conformal symmetry alone does not constrain the form of the function $g(u,v)$. 
Next we would like to perform the operator state mapping and take $z_3\rightarrow\infty$ and $z_4\rightarrow 0$ 
to absorb the $\mathcal{O}_2$ and $\mathcal{O}_2^{\dagger}$ operators into the initial and final states.
Using the transformation law (\ref{eq:cylindermap1}) of the correlator together with (\ref{eq:finalstate}) leads to
\beq
g(u,v)=\frac{\langle\mathcal{O}_2|\mathcal{O}_1(\tau_1,\phi_1)\mathcal{O}_1^{\dagger}(\tau_2,\phi_2)|\mathcal{O}_2\rangle}
{\langle0|\mathcal{O}_1(\tau_1,\phi_1)\mathcal{O}_1^{\dagger}(\tau_2,\phi_2)|0\rangle},\label{eq:4point}
\eeq
with
\beq
v=e^{2(\tau_1-\tau_2)},\quad u=1- 2 e^{\tau_1-\tau_2}\cos(\phi_1-\phi_2)+e^{2(\tau_1-\tau_2)}.
\eeq
Thus, we have reduced the full 4-point function into a two point function in the cylinder, in the state $|\mathcal{O}_2\rangle$.

\begin{figure}[h]
\begin{center}
\includegraphics[scale=.6]{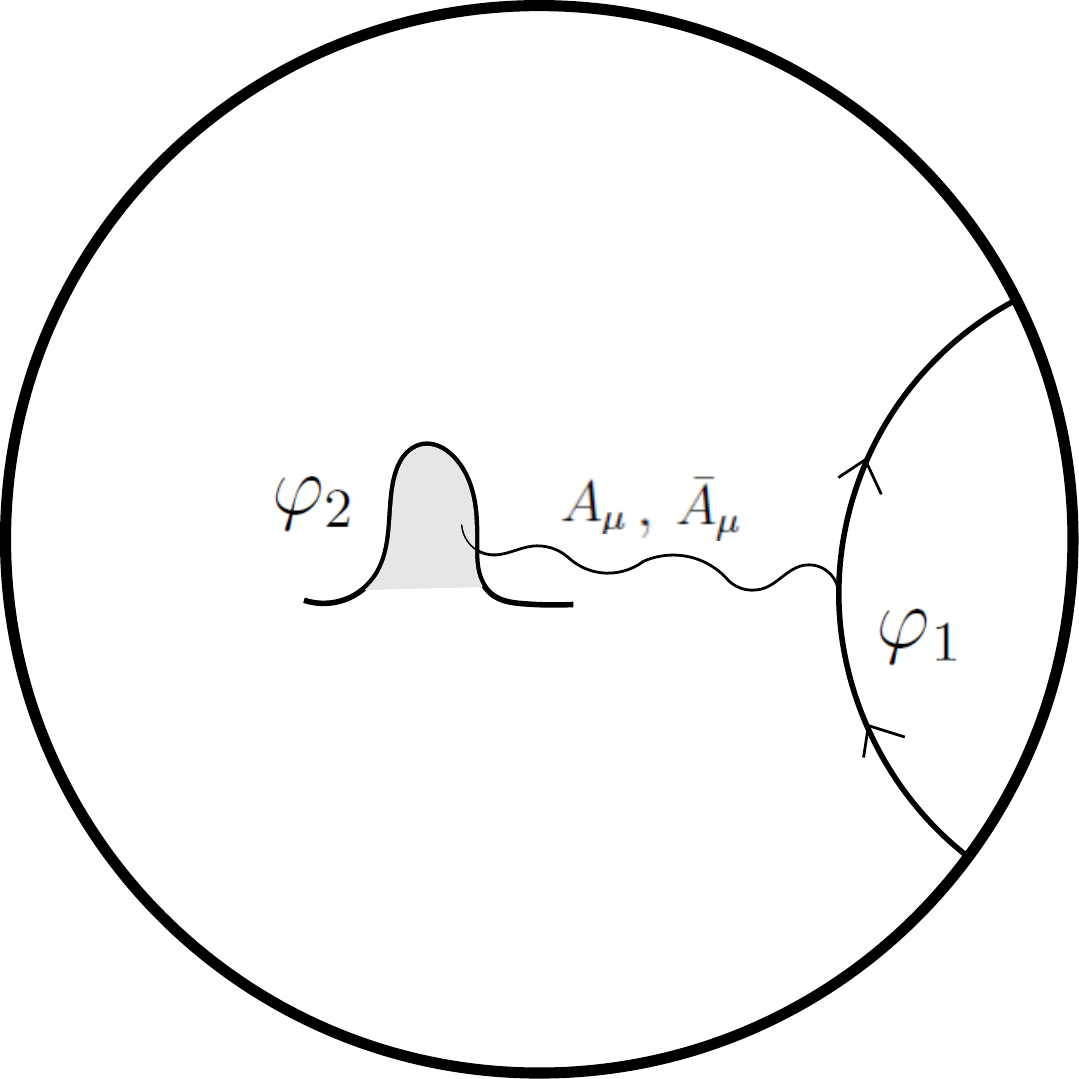}
\caption{\label{fig:diagram} The operator state mapping relates the four poing function into a two point function of $\varphi_1$
in the background of the $\varphi_2$ ground state. The grey blob represents the $\varphi_2$ ground state wavefunction. }
\end{center}
\end{figure}

For practical purposes we will again work in real time. In the bulk the procedure of the operator state mapping means that we prepare $\varphi_2$ into its ground state and compute the two point function of $\varphi_1$ in this background. This is visualized in Figure \ref{fig:diagram}.
When the Chern-Simons interaction is the only interaction between the fields, we are lead to calculate the $\varphi_1$ two point
function in the background Chern-Simons fields given by (\ref{eq:A1}) and (\ref{eq:A2}).\footnote{One should note that this prescription actually
leaves out disconnected self-energy contributions to the four point function. We will ignore these contribution everywhere in 
this manuscript.} The final simplification happens due to the fact
that $\varphi_1$ couples only to the combination $A+\bar{A}$. This combination is simply a constant
\beq
A+\bar{A}=\frac{2e_2}{k}dt=2\mu dt.
\eeq
This indeed follows directly from adding together the equations (\ref{eq:CSsourced}) and (\ref{eq:CSsourced2}), leading to an unsourced
Chern-Simons equation d$(A+\bar{A})=0$.

The overall effect of the state $|\mathcal{O}_2\rangle$ is thus the following. It introduces a constant background chemical potential $\mu=e_2/k$
for the scalar field $\varphi_1$. 
Now the beauty of this is that we do not even have to calculate the $\varphi_1$ two point function to find the full 4-point function!
To obtain the two point function we should find the complete set of solutions to the equation\footnote{Or alternatively the bulk to boundary propagator.}
\beq
\frac{1}{\sqrt{-g}}\partial_i(\sqrt{-g}g^{ij}\partial_j\varphi_1)+g^{tt}(\partial_t-ie_1\mu)^2\varphi_1-m_1^2\varphi_1=0.\label{eq:freescalar}
\eeq
Let us define a new field $\varphi'$ through $\varphi_1=e^{ie_1\mu t}\varphi'$. Plugging this into (\ref{eq:freescalar}) leads to a free Klein-Gordon equation for $\varphi'$ without a chemical potential. So the effect of the Chern-Simons
interaction is to shift the energies of all states of the $\varphi_1$ particle in the bulk by a constant $e_1\mu$. 
Thus, the bulk two point function of $\varphi_1$ is given by
\beq
e^{ie_1\mu(t_1-t_2)}\langle 0|\varphi'(x_1)\varphi'(x_2)^{\dagger}|0\rangle.
\eeq
The boundary two point function can be obtained as the boundary limit of the bulk two point function. 
Continuing back to Euclidean time $\tau=i t$ and plugging this to the formula for the 4-point function (\ref{eq:4point}) we see that everything else cancels out except
\beq
g(u,v)=e^{e_1\mu (\tau_1-\tau_2)}=v^{e_1\mu/2}.
\eeq
Since we are working in the first order in an expansion in $1/k$, we are only allowed to keep the first order in an expansion in $\mu$
\beq
g(u,v)=1+\frac{e_1e_2}{2k}\log v+O(1/k^2).
\eeq
Thus,to first order in $1/k$ the boundary four point function is given by
\beq
G_4=\frac{1}{|z_{12}|^{2\Delta_1}|z_{23}|^{2\Delta_2}}\Big(1+\frac{e_1e_2}{2k}\log v+O(1/k^2)\Big).\label{eq:scalar4point}
\eeq

\section{The CFT side of the story}\label{sec:crossing2}

The CFTs dual to weakly coupled bulk theories exhibit factorization of correlation functions of appropriately defined operators.
In the case of matrix field theories, the operators are appropriately normalized single trace operators.
Here we will assume a CFT that has at least two (single trace) operators $\mathcal{O}_1$ and $\mathcal{O}_2$, with corresponding scaling dimensions $\Delta_1$ and $\Delta_2$.

It is useful to organize the boundary theory four point function in terms of the operator product expansion. 
Operator product expansion is a general property of local quantum field theories. Consider the correlation function
\beq
\langle \mathcal{O}_i(0)\mathcal{O}_j(x)...\rangle,
\eeq
where the dots denote possible other operators. Operator product expansion states
that in the limit $x\rightarrow 0$, the above correlation function can be expanded in a series of local operators $\mathcal{O}_n(x)$ in the form
\beq
\langle \mathcal{O}_i(0)\mathcal{O}_j(x)...\rangle= \sum_n f_{ijn}(x)\langle\mathcal{O}_n(x)...\rangle,
\eeq
Conformal symmetry fixes the $x$ dependence of $f_{ijn}(x)$ into a power law. Furthermore, in a conformal field theory the operators $\mathcal{O}_n(x)$ can
be classified into conformal primary operators and their descendants (obtained by acting on the primaries with Virasoro generators).
Thus, the operator product expansion in a CFT can be organized in terms of a sum over conformal primaries only, with more complicated coefficient functions $f_{ijn}$ that include derivative operators.

The OPE can be applied twice to the four point function $G_4$ as $|z_{12}|\rightarrow 0$ and $|z_{34}|\rightarrow 0$. This reduces the four point function into
a double sum over primary operators
\beq
G_4=\sum_{n,m}D_{11^{\dagger}n}(x_{12},\partial_{1})D_{2^{\dagger}2m}(x_{34},\partial_{3})\langle\mathcal{O}_n(x_1)\mathcal{O}_m(x_3)\rangle.
\eeq
The derivative operators $D_{ijn}$ are fixed by conformal symmetry up to an overall normalization, which is the OPE coefficient $C_{ijn}$.
The double sum can be reduced into a single sum due to the orthogonality of the primary two point functions. 
The contribution of each primary operator within the sum is called a conformal block of the corresponding operator.

Let us apply this to our setting. We are assuming that we have a theory that has the scalar operators and
in addition at least the unit operator $1$ and the current operator $j^{\mu}$. Here we will show that the presence of 
these operators is enough to reproduce the physics of scalar fields coupled to Chern-Simons theory in the bulk. 
The following discussion still applies to more general CFT's, in which case the other operators give additional contributions
to the 4 point function. 

The OPE in this case reads
\beq
\mathcal{O}_i^{\dagger}(z_1)\mathcal{O}_i(z_2)=\frac{1}{|z_{12}|^{2\Delta_i}}\Big(1+
\sum_{n=0}^{\infty}c^i_nz_{12}^{n+1}\partial_{z_2}^n j_z(z_2)+h.c.+...\Big),\label{eq:currentope1}
\eeq
where the first term corresponds to the unit operator and the sum corresponds to the current operator and its descendants. The OPE coefficients are worked out in Appendix \ref{sec:opecoeff} and are given by
\beq
c^i_n=\frac{C_{\mathcal{O}_i^{\dagger}j\mathcal{O}_i}}{(n+1)!C_V}
\eeq
Above we have only included the global conformal descendants of the current operator,
while in a 1+1 CFT, as the conformal symmetry algebra gets enhanced to the Virasoro algebra, there are also Virasoro descendants
of the current operator present. The reason for neglecting these operators from the OPE is that their contributions are suppressed at large central
charge. A more detailed discussion is included in Appendix \ref{sec:virasoro}. 

Using the OPE (\ref{eq:currentope1}) twice within the 4 point function (\ref{eq:4pointdef}) gives
\beq
G_4=\frac{1}{|z_{12}|^{2\Delta_1}|z_{34}|^{2\Delta_2}}\Big(1+\sum_{n,m}(c_n^1)^*c_m^2z_{12}^{n+1}z_{34}^{m+1}\partial_{z_2}^n
\partial_{z_4}^m\langle j_z(z_2)j_z(z_4)\rangle+c.c.\Big).
\eeq
Thus, we should perform the double sum
\beq
S=\sum_{n,m}\frac{z_{12}^{n+1}z_{34}^{m+1}}{(n+1)!(m+1)!}\partial_{z_2}^n\partial_{z_4}^m\frac{1}{z^2}=\Big(\sum_{n=0}^{\infty}
\frac{z_{12}^{n+1}}{(n+1)!}\partial_{z_2}^n\Big)\Big(\sum_{m=0}^{\infty}
\frac{z_{34}^{m+1}}{(m+1)!}\partial_{z_4}^n\frac{1}{z_{24}^2}\Big)
\eeq
Using the identity $\partial^n(1/z^2)=(-1)^n(n+1)!z^{-n-2}$ we can perform the sum over $m$ first to give
\beq
S=\sum_{n=0}^{\infty}
\frac{z_{12}^{n+1}}{(n+1)!}\partial_{z_2}^n\Big(\frac{1}{z_{23}}-\frac{1}{z_{24}}\Big)
\eeq
Then, using $\partial_z^n(1/z)=(-1)^nn!z^{-n-1}$ we get
\beq
S=\log\Big(\frac{z_{13}z_{24}}{z_{23}z_{14}}\Big).
\eeq
Thus, the current operator and its descendants give the following contribution to the four point function 
\beq
G_4=\frac{1}{|z_{12}|^{2\Delta_1}|z_{34}|^{2\Delta_2}}\Big(1+\frac{C_{\mathcal{O}_1^{\dagger}j\mathcal{O}_1}^*C_{\mathcal{O}_2^{\dagger}j\mathcal{O}_2}}{2C_V}\log v+...\Big),
\eeq
where the dots denote possible other operators in the CFT. The function $\log v$ is the conformal block of the current
operator\cite{Dolan:2000ut} . Plugging in the values of the OPE coefficients and of $C_V$, gives 
\beq
G_4=\frac{1}{|z_{12}|^{2\Delta_1}|z_{34}|^{2\Delta_2}}\Big(1+\frac{e_1e_2}{2k}\log v\Big).\label{eq:G42}
\eeq
This indeed is exactly what we found from bulk Chern-Simons theory.

\section{A simple check of the results}\label{sec:check}

As a simple check of the results for the 4-point functions we show how it reproduces the
binding energies of charged particles in the bulk. From the perspective of the CFT on a plane, the binding energy of 
two bulk particles $\varphi_1$ and $\varphi_2$ is given by the difference of the scaling dimension of the composite 
operator $\mathcal{O}_1\mathcal{O}_2$ from the free result $\Delta_1+\Delta_2$. To obtain
this scaling dimension we consider the two point function of composite operators
\beq
G_2=\langle (\mathcal{O}_1\mathcal{O}_2)(z,\bar{z}) (\mathcal{O}_1\mathcal{O}_2)^{\dagger}(z',\bar{z}')\rangle.
\eeq
This can be obtained from $G_4$ as a limit\footnote{Strictly speaking we should introduce a multiplicative 
renormalization factor relating $(\mathcal{O}_1\mathcal{O}_2)(x)=\lim_{\epsilon\rightarrow 0}Z(\epsilon)
\mathcal{O}_1(x+\epsilon)\mathcal{O}_2(x)$, to have a finite two point function in the limit $\epsilon\rightarrow 0$.} where $|z_{14}|=|z_{23}|=\epsilon\rightarrow 0$,
while $|z_{12}|\approx |z-z'|$. In this limit, after using
$v\approx \epsilon^4/|z-z'|^4$, the four point function becomes
\beq
G_4=\frac{1}{|z-z'|^{2\Delta_1+2\Delta_2}}\Big(1-\frac{e_1e_2}{k}\log\frac{|z-z'|^2}{\epsilon^2}+...\Big),\label{eq:G422}
\eeq
The logarithmic term can be recognized as a first term in the series of the composite operator two point function
\beq
\frac{1}{|z-z'|^{2(\Delta_1+\Delta_2+\gamma)}},
\eeq
in powers of the anomalous dimension $\gamma$. Thus, we can identify
\beq
\gamma=\frac{e_1e_2}{k}.\label{eq:anomalousCFT}
\eeq

Next we consider the binding energy from the bulk point of view. We start by considering classical charged particles in the bulk.
A classical particle with charge $e$ at the center of AdS$_3$ gives rise to a current
\beq
\sqrt{-g}J^{\mu}=e\frac{dx^{\mu}}{dt}\delta^{(2)}(x).
\eeq
The Chern-Simons equations of motion then give
\beq
F_{\theta\phi}=-\frac{2\pi e}{k}\delta^{(2)}(x),\quad \bar{F}_{\theta\phi}=\frac{2\pi e}{k}\delta^{(2)}(x).
\eeq
Thus, the holonomies of the gauge fields around a constant $\theta$ circle are given by
\beq
W_A=e^{i\oint A}= e^{i\int F}=e^{\frac{2i\pi e}{k}},\quad W_{\Bar{A}}=e^{i\int \bar{F}}=e^{-\frac{2i\pi e}{k}}.
\eeq
In the gauge $A_{\theta}=0$ this implies $A_{\phi}=\frac{e}{k}$ and $\bar{A}_{\phi}=-\frac{e}{k}$ outside the
center of AdS$_3$ where the gauge field is flat. The time components are fixed by the boundary conditions at the
AdS boundary. Thus, the gauge fields due to a classical point particle are given by
\beq
A=\frac{e}{k}(dt+d\phi),\quad \bar{A}=\frac{e}{k}(dt-d\phi).\label{eq:asymptoticfields}
\eeq

The gauge field contribution to the
holographic energy momentum tensor separates from the rest of the energy momentum tensor since the Chern-Simons action is independent of the spacetime
metric. Thus, the Chern-Simons contribution to the holographic energy momentum tensor appears from a boundary term, and is given by\footnote{We assume below that $k$ is positive} \cite{Jensen:2010em}
\beq
T_{\mu\nu}=\frac{k}{8\pi}\Big(A_{\mu}A_{\nu}+\bar{A}_{\mu}\bar{A}_{\nu}-A_{\mu}\bar{A}_{\nu}-A_{\nu}\bar{A}_{\mu}-\frac{1}{2}\eta_{\mu\nu}(A^2+\bar{A}^2-2 A\cdot\bar{A})\Big).\label{eq:energymom}
\eeq
This gives us the Chern-Simons contribution to the energy of the particle as
\beq
E_{CS}(e)=\int_0^{2\pi}T_{tt}=\frac{e^2}{2k}.
\eeq
Two particles with charges $e_1$ and $e_2$ give rise to the same holonomy as a single particle with charge $e_1+e_2$. Thus,
the gauge fields near the boundary have the same form (\ref{eq:asymptoticfields}) with $e=e_1+e_2$. Due to the non-linearity
of the holographic energy momentum tensor of Chern-Simons theory, there is an interaction energy between two charged particles
\beq
\delta E=E_{CS}(e_1+e_2)-E_{CS}(e_1)-E_{CS}(e_2)=\frac{e_1e_2}{k},
\eeq
This is indeed the anomalous dimension we obtain from the 4-point function.

\section{Discussion}

The effect of bulk Chern-Simons fields in the scalar four point function is reproduced in the leading order in $1/k$ by the operator product 
expansion of the operators $1$, and $j_{\mu}$ and its (global) conformal descendants. Since the CFT under consideration
must also include the scalar operators $\mathcal{O}_1$ and $\mathcal{O}_2$, we could have very well added operators of the
form
\beq
\mathcal{O}_i^{\dagger}\overleftrightarrow{\Box}^n\overleftrightarrow{\partial}_{\mu_1}...\overleftrightarrow{\partial}_{\mu_l}\mathcal{O}_i-\textrm{traces}\label{eq:composite1}
\eeq
to the OPE. Indeed in order to satisfy crossing symmetry of the OPE one must include operators of the form 
\beq
\mathcal{O}_1 \overleftrightarrow{\Box}^n\overleftrightarrow{\partial}_{\mu_1}...\overleftrightarrow{\partial}_{\mu_l}\mathcal{O}_2-\textrm{traces},\label{eq:composite2}
\eeq
in the $|z_{14}|\rightarrow 0$ OPE channel, as discussed in Appendix \ref{sec:crossingsym}. Thus, from the point
of view of a generic CFT there is no reason for leaving out operators of the form (\ref{eq:composite1}) from the $|z_{12}|\rightarrow 0$ OPE.
Including the operators (\ref{eq:composite1}) would lead to an infinite number of possible four point functions. These four point functions must satisfy
crossing symmetry, which gives an infinite number of constraints still leading to an infinite number of possible four point
functions. 

Pure Chern-Simons interaction between the scalars thus corresponds to a particularly minimal operator content in the OPE.
Conformal blocks of most CFT operators are not crossing symmetric by themselves\footnote{Meaning with any operator content in the other $z_{14}\rightarrow 0$ channel.} and thus cannot give rise to a consistent
four point function without the addition of an infinite set of further operators \cite{ElShowk:2011ag}. Thus, most
bulk particle exchange interactions cannot be directly interpreted as due to the conformal block of the corresponding 
operator dual to the bulk field being exchanged. The key simplification in the case of the current operator
in 1+1 dimensions is that its conformal block is crossing symmetric (with an appropriate choice of composite operators (\ref{eq:composite2}) in
the other $z_{14}\rightarrow 0$ channel). It is straightforward to demonstrate that the current conformal block
in 1+1 dimensions is crossing symmetric. The details are included in Appendix \ref{sec:crossingsym}. 

As shown in Appendix \ref{sec:crossingsym} one can read off the anomalous dimensions of all the operators of the
form (\ref{eq:composite2}) from the four point function (\ref{eq:G42})
\beq
\gamma(n,l)=\frac{e_1e_2}{k}.\label{eq:anomalousnk}
\eeq
As discussed in \cite{Fitzpatrick:2012yx}, the $l$ dependence of the anomalous dimension is
related to how the corresponding bulk interaction varies with bulk (proper) distance. This follows
as at sufficiently large $l$ the distance between the bulk fields in the state dual to the operator
(\ref{eq:composite2}) scales as  $b\propto \log l$.
If we had started from the CFT side we could ask what kind of a bulk interaction could
reproduce the contribution of the current conformal block. Knowing the 
anomalous dimensions (\ref{eq:anomalousnk}) from the block, we would be lead to the conclusion that
whatever bulk interaction is dual to it, the interaction has to
be independent of the bulk distance between the particles.\footnote{Strictly speaking this
can be concluded only for large $l$, but even for smaller $l$ even thought no obvious quantitative
connection between bulk distance and $l$ exists, it still seems likely that the only way to get
anomalous dimensions independent of $l$ and $n$ is to have an interaction independent of distance.}
This is of course consistent with Chern-Simons interaction, which as we saw in section \ref{sec:4point},
has no local bulk effects but simply shifts the energies of all charged bulk particle states by a constant.

\section*{Aknowledgements}

I would like to thank A. O' Bannon, A. Starinets and K.P. Yogendran for discussions. This research was supported by
the European Research Council under the European Union’s Seventh Framework Programme (ERC Grant
agreement 307955).

\begin{appendix}

\section{Some details on conventions}

Integral measures $d^3x$ are always defined to be positive.
Complex coordinates have been defined through
\beq
z=x+i t_E,\quad w=\tau+i\phi.
\eeq
These may be continued to real time using $t_L=-it_E$ and $t=-i\tau$. Throughout we assume (without loss
of generality) that the level $k$ is positive. All the $\epsilon$ symbols we use are defined as
\beq
\epsilon^{uz\bar{z}}=1,\quad \epsilon^{uxt_L}=1,\quad \epsilon^{t\theta\phi}=1.
\eeq
The boundary terms are evaluated at $u=\epsilon$, where the induced metric is $\gamma_{z\bar{z}}=1/(2\epsilon^2)$, $\gamma^{z\bar{z}}=2\epsilon^2$ and $\sqrt{|\gamma|}=1/(2\epsilon^2)$.

\section{Current OPE coefficient}\label{sec:C}

In this Appendix we review how the $\mathcal{O}^{\dagger}j\mathcal{O}$ OPE coefficient is fixed
by a current conservation Ward identity in the CFT. The OPE coefficient is defined
through the 3-point function
\beq
\langle \mathcal{O}^{\dagger}(x_1)j^{\mu}(x_2)\mathcal{O}(x_3)\rangle=C_{\mathcal{O}^{\dagger}j\mathcal{O}}\frac{Z^{\mu}}{|x_1-x_3|^{2\Delta_i}},\quad Z^{\mu}=\frac{x_2^{\mu}-x_3^{\mu}}{(x_2-x_3)^2}-\frac{x_2^{\mu}-x_1^{\mu}}{(x_2-x_1)^2}.\label{eq:3point}
\eeq
Current conservation Ward identities tell us that
\beq
\langle \mathcal{O}^{\dagger}(x_1)\partial_{\mu}j^{\mu}(x_2)\mathcal{O}(x_3)\rangle=-ie(\delta^{(2)}(x_2-x_3)-\delta^{(2)}(x_2-x_1))\langle\mathcal{O}^{\dagger}(x_1)\mathcal{O}(x_2)\rangle.\label{eq:Ward}
\eeq
This Ward identity can be derived for example using standard path integral methods \cite{DiFrancesco:1997nk} and we have used the fact the $\mathcal{O}$ transforms as $\mathcal{O}\rightarrow e^{ie\alpha}\mathcal{O}$
under the global symmetry transformations. Requiring (\ref{eq:3point}) to be consistent with (\ref{eq:Ward}) fixes the OPE coefficient $C_{\mathcal{O}^{\dagger}j\mathcal{O}}$ since
\beq
\partial_{\mu,x_2}\langle \mathcal{O}^{\dagger}(x_1)j^{\mu}(x_2)\mathcal{O}(x_3)\rangle=-2\pi C_{\mathcal{O}^{\dagger}j\mathcal{O}}\frac{1}{|x_1-x_2|^{2\Delta_i}}(\delta^{(2)}(x_2-x_3)-\delta^{(2)}(x_2-x_1)),
\eeq
where we used $\partial_{\mu,x}(x^{\mu}-y^{\mu})/(x-y)^2=-2\pi\delta^{(2)}(x-y)$. Comparing to the Ward identity gives
\beq
C_{\mathcal{O}^{\dagger}j\mathcal{O}}=i\frac{e}{2\pi}.
\eeq
The OPE coefficient is symmetric under the change of its indices, except when changing the order of $\mathcal{O}^{\dagger}$ and
$\mathcal{O}$, under which the OPE coefficient changes sign
\beq
C_{\mathcal{O}j\mathcal{O}^{\dagger}}=-C_{\mathcal{O}^{\dagger}j\mathcal{O}}=-i\frac{e}{2\pi}
\eeq

\section{OPE coefficients between the scalars and the current operator}\label{sec:opecoeff}

The operator product expansion of a charged scalar operator $\mathcal{O}$ in a generic 1+1 CFT includes the contributions of the current operator
\beq
\mathcal{O}^{\dagger}(z_1)\mathcal{O}(z_3)=\frac{1}{|z_{13}|^{2\Delta}}\sum_n c_n z_{13}^{n+1}\partial^nj_z(z_3)+...,\label{eq:opeapp}
\eeq
where the dots correspond to other operators in the theory and include the same sum for the right moving component of the 
current $j_{\bar{z}}$. Since the analysis for the right moving part is identical to the one for the left moving component, we will
only discuss the latter here in detail. Also (\ref{eq:opeapp}) only includes the global conformal descendants $L_{-1}^k j_z$.
The reason for leaving out the higher Virasoro descendants $L_{-n}^k$ is that they give rise to subleading contributions in the limit
of large central charge. This is explained in more detail in the Appendix \ref{sec:virasoro}. 

To fix the unknown coefficients $c_n$, we compare the full three point function
\beq
\langle \mathcal{O}^{\dagger}(z_1)j_z(z_2)\mathcal{O}(z_3)\rangle=\frac{C_{\mathcal{O}^{\dagger}j\mathcal{O}}}{2}\frac{1}{|z_{13}|^{2\Delta}}
\Big(\frac{1}{z_{23}}-\frac{1}{z_{21}}\Big),\label{eq:3pointapp}
\eeq
to the one obtained from operator product expansion
\beq
\langle \mathcal{O}^{\dagger}(z_1)j_z(z_2)\mathcal{O}(z_3)\rangle=
\frac{1}{|z_{13}|^{2\Delta}}\sum_n c_n z_{13}^{n+1}\langle \partial^nj_z(z_3) j_z(z_2)\rangle.
\eeq
Using the known current two point function gives
\beq
\langle \mathcal{O}^{\dagger}(z_1)j_z(z_2)\mathcal{O}(z_3)\rangle=
-\frac{C_V}{2}\frac{1}{|z_{13}|^{2\Delta}}\sum_n c_n (n+1)! z_{13}^{n+1}z_{23}^{-n-2}.
\eeq
On the other hand series expanding (\ref{eq:3pointapp}) gives
\beq
\langle \mathcal{O}^{\dagger}(z_1)j_z(z_2)\mathcal{O}(z_3)\rangle=-\frac{C_{\mathcal{O}^{\dagger}j\mathcal{O}}}{2}\frac{1}{|z_{13}|^{2\Delta}}
\sum_n z_{13}^{n+1}z_{23}^{-n-2}.
\eeq
Matching the coefficients of the two series fixes the OPE coefficients
\beq
c_n=\frac{C_{\mathcal{O}^{\dagger}j\mathcal{O}}}{(n+1)!C_V}.
\eeq

\section{Decoupling of Virasoro descendants at large $c$}\label{sec:virasoro}

This can be straightforwardly seen from the Virasoro algebra
\beq
\[L_{n},L_{m}\]=(n-m)L_{n+m}+\frac{c}{2}n(n^2-1)\delta_{n+m,0}.
\eeq
As an example consider the descendant $L_{-n}j_z$. The two point function of this operator is proportional to the norm
\beq
\langle L_{-n}j_z|L_{-n}j_z\rangle=\langle j_z|L_nL_{-n}|j_z\rangle=(\frac{c}{2}n(n^2-1)+2n)\langle j_z|j_z\rangle.
\eeq
Thus, for $n>1$, the two point function scales as $c^1$ as $c\rightarrow\infty$. 

On the other hand the three point functions
\beq
\langle \mathcal{O}^{\dagger}(z_1)L_{-n}j_z(z_2)\mathcal{O}(z_3)\rangle\label{eq:3pointapp2}
\eeq
are obtained from (\ref{eq:3pointapp}), by acting with derivative operators \cite{DiFrancesco:1997nk}. Thus, as far as counting
powers of $c$ goes, these three point functions are order $c^0$. Using the same procedure as
in the previous appendix to obtain the OPE coefficients of the higher descendants by considering (\ref{eq:3pointapp2}),
we find that the OPE coefficients are of the order $c^{-1}$. Thus, we can ignore them in the leading order at
large $c$.

\section{On crossing symmetry}\label{sec:crossingsym}

We have not yet shown whether the operator content we took into account is consistent, in the sense that it should satisfy crossing symmetry.
This is the purpose of this Appendix. 

\begin{figure}[h]
\begin{center}
\includegraphics[scale=.4]{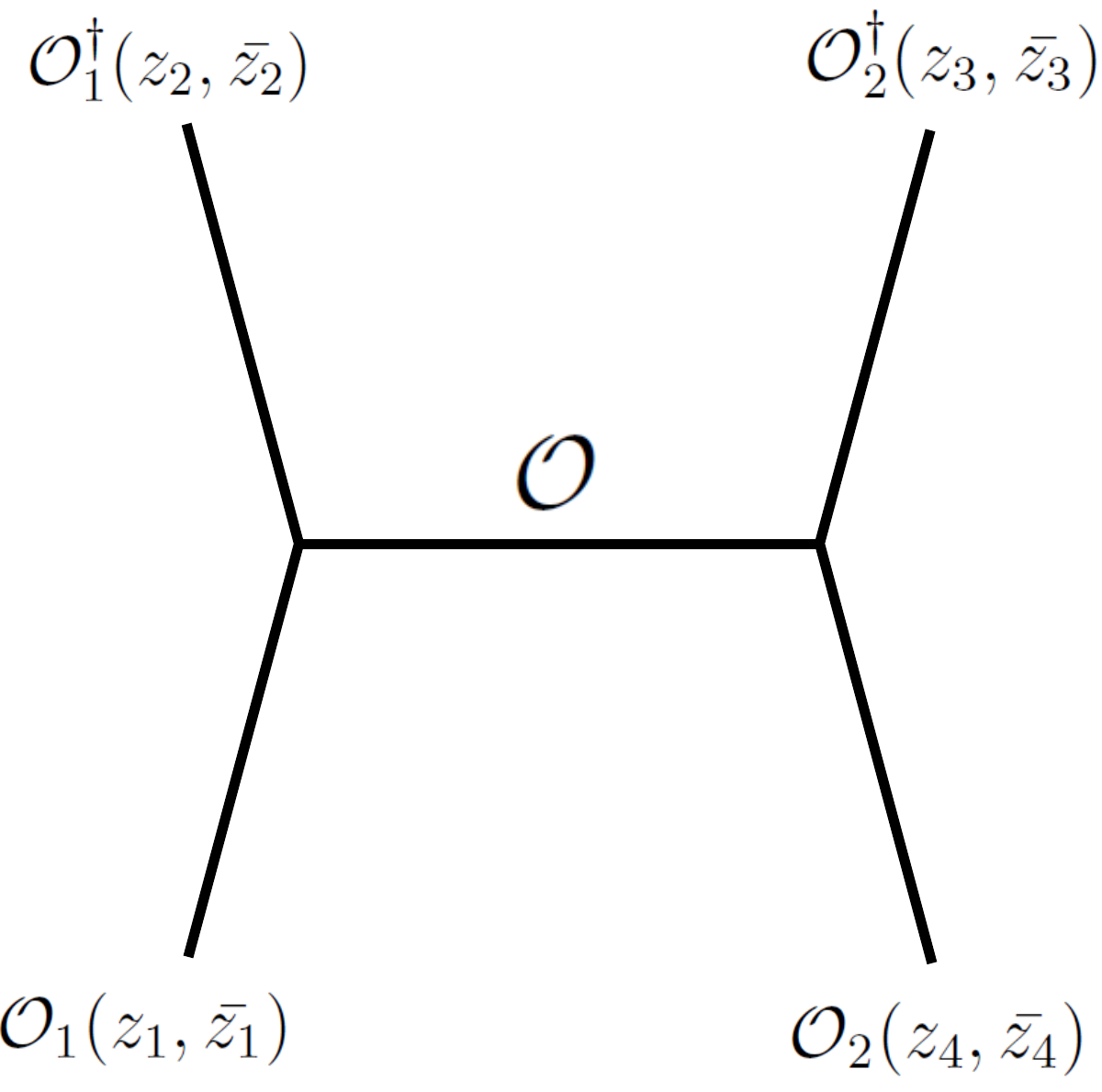}
\quad\quad
\includegraphics[scale=.4]{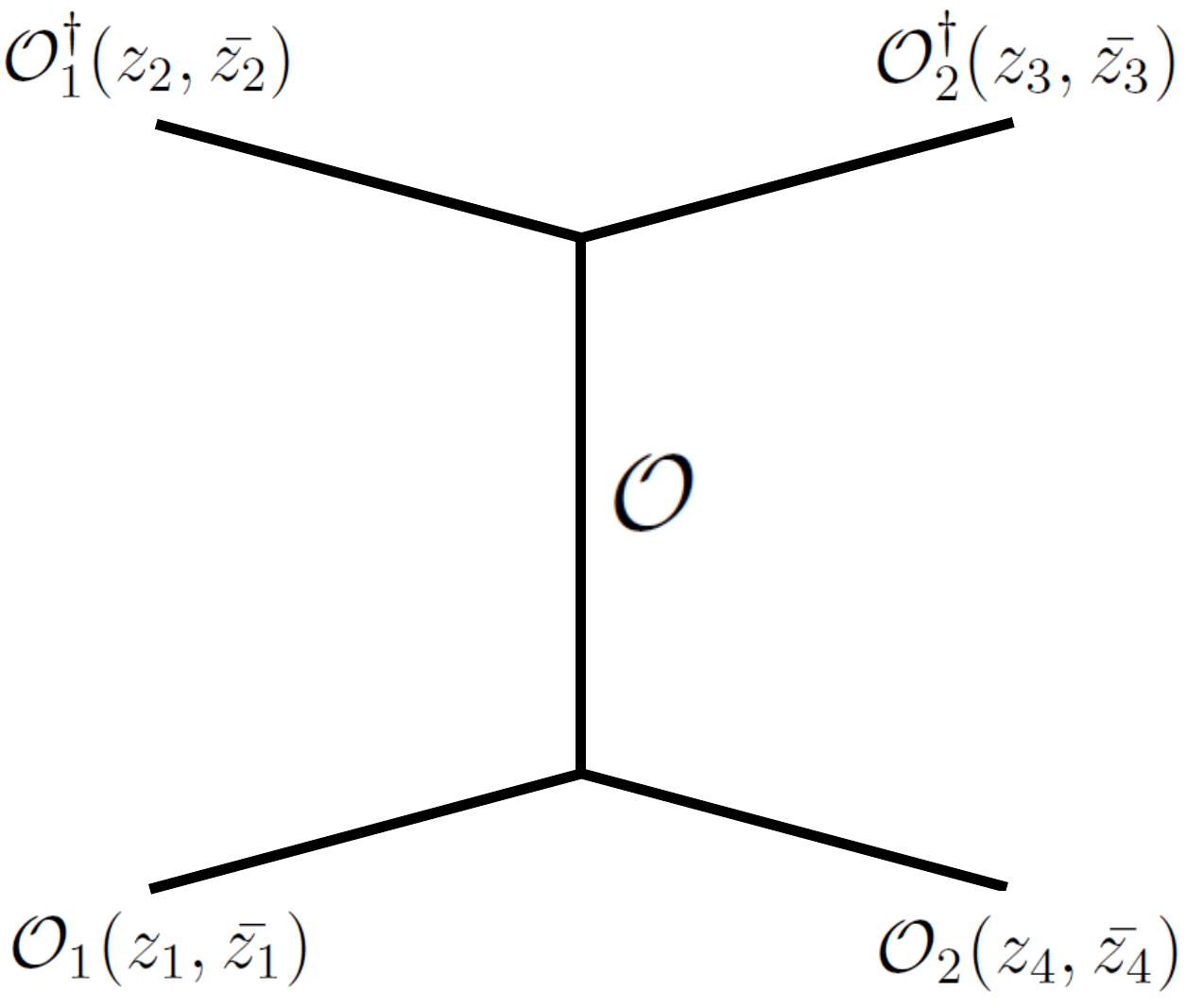}
\caption{\label{fig:ope} Left: OPE in the S-channel. Right: OPE in the T-channel. The OPE sums over the possible operators $\mathcal{O}$. }
\end{center}
\end{figure}

Using the decomposition of the 4 point function in (\ref{eq:F}), the OPE can be organized into primary operators
\beq
g(u,v)=\sum_{\mathcal{O}}C_{11\mathcal{O}}C_{22\mathcal{O}}g_{\Delta,l}(\eta,\bar{\eta}),\label{eq:Schannel}
\eeq
where $\Delta$ and $l$ are the scaling dimension and the spin of the primary operator and
\beq
u=\eta\bar{\eta},\quad v=(1-\eta)(1-\bar{\eta}).
\eeq
The function $g_{\Delta,l}(\eta,\bar{\eta})$ is given by \cite{Dolan:2000ut}
\beq
g_{\Delta,l}(\eta,\bar{\eta})=\Big(-\frac{1}{2}\Big)^l 2^{-\delta_{l0}}\Big(k(\Delta+l,\eta)k(\Delta-l,\bar{\eta})+k(\Delta+l,\bar{\eta})k(\Delta-l,\eta)\Big),\label{eq:block}
\eeq
where we used the notation\cite{Heemskerk:2009pn}
\beq
k(\beta,\eta)=\eta^{\beta/2}\,_2F_1(\beta/2,\beta/2,\beta,\eta).
\eeq
This is the operator product expansion in the S-channel (see Figure \ref{fig:ope}). We can also perform the OPE in the region $|z_{14}|\rightarrow 0$ and $|z_{23}|\rightarrow 0$. This is called the T-channel OPE and is given in terms of the conformal blocks as
\beq
g(u,v)=\sum_{\mathcal{O}}C_{12\mathcal{O}}C_{1^{\dagger}2^{\dagger}\mathcal{O}^{\dagger}}f_{\Delta,l}(\eta,\bar{\eta}),\label{eq:Schannel}
\eeq
where
\beq
f_{\Delta,l}(\eta,\bar{\eta})=(-1)^l(\eta\bar{\eta})^{\Delta_2}((1-\eta)(1-\bar{\eta}))^{\Delta-\Delta_1-\Delta_2-l}F(\Delta,l),
\eeq
and
\beq
F(\Delta,l)=\frac{1}{2^l 2^{-\delta_{l0}}}\Big((1-\eta)^l\,_2F_1(a,a,\Delta+l,1-\eta)\,_2F_1(a-l,a-l,\Delta-l,1-\bar{\eta})+(\eta\leftrightarrow \bar{\eta})\Big),\nonumber
\eeq
where $a=(\Delta+\Delta_2-\Delta_1+l)/2$.

Crossing symmetry is the requirement that the four point function should be independent of the channel one chooses to expand on. As an equation, this reads
\beq
\sum_{\mathcal{O}}C_{11^{\dagger}\mathcal{O}}C_{2^{\dagger}2\mathcal{O}}g_{\Delta,l}(\eta,\bar{\eta})=\sum_{\mathcal{O}}C_{12\mathcal{O}}C_{1^{\dagger}2^{\dagger}\mathcal{O}^{\dagger}}f_{\Delta,l}(\eta,\bar{\eta})\label{eq:crossing}.
\eeq
With the operator content $(1,J_{\mu})$ the S-channel OPE is simply given by
\beq
g(u,v)=1+\frac{1}{2}\frac{e_1e_2}{k}\log v.
\eeq
The T-channel OPE involves the primary operators of the schematic form $\mathcal{O}_1(\partial^2)^n\partial_1...\partial_l\mathcal{O}_2$, which have
scaling dimensions 
\beq
\Delta(n,l)=\Delta_1+\Delta_2+2n+l+\gamma(n,l)=\Delta_0(n,l)+\gamma(n,l),
\eeq
where $\gamma(n,l)$ are the corresponding anomalous dimensions assumed to be of the order $1/k$.
Furthermore we will denote the OPE coefficients as $c(n,l)=(-1)^l C_{12\mathcal{O}}C_{1^{\dagger}2^{\dagger}\mathcal{O}^{\dagger}}=|C_{12\mathcal{O}}|^2$. Defined
this way, the coefficients $c(n,l)$ must be positive. Thus, the crossing symmetry equation
takes the form of a double sum
\beq
\sum_{n,l}c(n,l)u^{\Delta_2}v^{n+\frac{1}{2}\gamma(n,l)}F(\Delta_0(n,l)+\gamma(n,l),l)=1+\frac{1}{2}\frac{e_1e_2}{k}\log v.\label{eq:fullcrossing}
\eeq
We would like to solve this equation order by order in $1/k$. Thus, we expand the OPE coefficients as $c(n,l)=c_0(n,l)+\delta c(n,l)$, where $\delta c(n,l)$
is of the order $1/k$.

\paragraph{Zeroth order:} To zeroth order in the $1/k$ expansion we get
\beq
\sum_{n,l}c_0(n,l)v^{n}F(\Delta_0,l)=u^{-\Delta_2}\label{eq:0thcrossing}
\eeq
This equation can be uniquely solved for the OPE coefficients $c_0(n,l)$. Let us define the variables $a=1-\eta$ and $b=1-\bar{\eta}$.
We will solve (\ref{eq:0thcrossing}) by series expanding around $a=0$ and $b=0$.

Using the hypergeometric series\footnote{We use the following convention for the Pochhammer symbol
$(x)_n=\Gamma(x+n)/\Gamma(x)$} $\,_2F_1(a,b,c,x)=\sum_{n=0}^{\infty}(a)_n(b)_n/(c)_n x^n/n!$ it is straightforward to expand (\ref{eq:0thcrossing})
as a power series in $a$ and $b$
\beq
\sum_{m,k} h_1(m,k) a^mb^k=\sum_{n,l,k,m} c_0(n,l) h_2(n,l,m,k)
\Big(a^{k+n}b^{m+l+n}+a^{m+l+n}b^{k+n}\Big),\label{eq:appseries}
\eeq
where all of the sums range from $0$ to $\infty$, and
\beq
h_1(m,k)=\frac{\Gamma(\Delta_2+m)\Gamma(\Delta_2+k)}{\Gamma(\Delta_2)^2m!k!},\quad  h_2(n,l,m,k)=\frac{(\alpha)^2_k(\beta)^2_m}{2^{l+\delta_{l0}}(\beta)_k(\delta)_mk!m!}.\label{eq:appcoeff}
\eeq
where $\alpha=\Delta_2+n+l$, $\beta=\Delta_1+\Delta_2+2n+2l$, $\gamma=\Delta_2+n$ and $\delta=\Delta_1+\Delta_2+2n$.
The equality in (\ref{eq:appseries}) should hold for each power of $a$ and $b$ separately. This leads us to the following method of solving (\ref{eq:appseries}) iteratively.
Consider first the powers $a^0b^j$ in (\ref{eq:appseries}), for some fixed $j$, which gives
\beq
h_1(0,j)=\sum_{l=0}^{j}h_2(0,l,j-l,0)c_0(0,l)+h_2(0,0,0,j)c_0(0,0).\label{eq:app0j}
\eeq
This can be solved iteratively to give $c_0(0,l)$ for all values of $l$. Taking $j=0$ gives $h_1(0,0)=2h_2(0,0,0,0)c_0(0,0)$, which after using (\ref{eq:appcoeff}) gives
$c_0(0,0)=1$. Next we can solve $c_0(0,1)$, $c_0(0,2)$ and so forth, up to arbitrarily high order. Thus, we can in principle determine all $c_0(0,j)$ from (\ref{eq:app0j}). For example the first
few terms are
\beq
c_0(0,0)=1,\quad c_0(0,1)=\frac{2\Delta_1\Delta_2}{\Delta_1+\Delta_2},\quad c_0(0,2)=\frac{2\Delta_1(\Delta_1+1)\Delta_2(\Delta_2+2)}{(1+\Delta_1+\Delta_2)(2+\Delta_1+\Delta_2)}.
\eeq
Now we can consider the $a^ib^j$ term in (\ref{eq:appseries}). The key is to note that on the right hand side the sum over $n$ gives non-vanishing contributions only when
$n=0,...,i$. This means that only the coefficients $c(n,l)$ with $n\le i$ and $l\le j$ appear in (\ref{eq:appseries}). Thus, we can again first take $i=1$ and solve the equations for all $j$ and then move to $i=2$ and solve for all $j$ and so on. Thus, in principle, all of the coefficients
$c_0(n,l)$ are uniquely determined by crossing symmetry. As so far we are considering a generalized free field theory, we will not pursue
the specific forms of $c_0(n,l)$, but will simply assume that the unique solution for the coefficients is consistent, i.e. $c_0(n,l)\ge 0$.

\paragraph{First order:} To first order in $1/k$ the crossing symmetry equation (\ref{eq:fullcrossing}) becomes
\begin{align}
&\sum_{n,l}\Big(\delta c(n,l)  v^n F(\Delta_0,l)+\frac{1}{2}c_0(n,l)\gamma(n,l)v^n F(\Delta_0,l)\log v\nonumber
\\
&+\frac{1}{2}c_0(n,l)\gamma(n,l)v^n\frac{\partial}{\partial n}F(\Delta_0,l)\Big)=\frac{1}{2}\frac{e_1e_2}{k}\log v,\label{eq:1stcrossing}
\end{align}
Identifying the logarithmic parts\footnote{This can be done because the $F(\Delta_0,l)$ does not have logarithmic terms near $v=0$.} gives
\beq
\sum_{n,l}v^nc_0(n,l)\gamma(n,l)F(\Delta_0,l)=u^{-\Delta_2}\frac{e_1e_2}{k}.
\eeq
Comparing this to (\ref{eq:0thcrossing}) immediately gives us the unique solution for the anomalous dimensions
\beq
\gamma(n,l)=\frac{e_1e_2}{k}.\label{eq:allanomalous}
\eeq
Thus, all of the composite higher spin operators have the same anomalous dimension. Finally we can solve for the $\delta c(n,l)$ coefficients from
the rest of (\ref{eq:1stcrossing})
\beq
\sum_{n,l}v^n\Big(\delta c(n,l)F(\Delta_0,l)+\frac{1}{2}c_0(n,l)\gamma(n,l)\frac{\partial}{\partial n}F(\Delta_0,l)\Big)=0.
\eeq
Using again $a=1-\eta$ and $b=1-\bar{\eta}$ and series expanding gives
\begin{align}
\sum_{n,l,k,m}\Big[\Big(\delta c(n,l)h_2(n,l,m,k)&+\frac{1}{2}\gamma(n,l)c(n,l)\partial_n h_2(n,l,m,k)\Big)\times\nonumber
\\
&\times(a^{k+n}b^{m+l+n}+a^{m+l+n}b^{k+n})\Big]=0,\label{eq:2ndcrossing}
\end{align}
We can use the same iterative procedure to uniquely solve the equation for $\delta c(n,l)$
as we used to solve for $c_0(n,l)$  as the equations have the same structure in terms of powers
of $a$ and $b$. The few lowest coefficients are given by
\begin{align}
&\delta c(0,0)=0,\quad \delta c(0,1)=-\frac{2e_1e_2\Delta_1\Delta_2}{k(\Delta_1+\Delta_2)^2},\nonumber
\\
&\delta c(0,2)=-\frac{2e_1e_2\Delta_1(\Delta_1+1)\Delta_2(\Delta_2+1)(3+2\Delta_2+2\Delta_1)}{k(\Delta_1+\Delta_2+1)^2(\Delta_1+\Delta_2+2)^2}.
\end{align}
This procedure is guaranteed to give a solution to all of the coefficients $\delta c(n,l)$ because the coefficients $h_2(n,l,m,k)$ in front of $\delta c(n,l)$ are all non-vanishing.Thus, we have demonstrated (up to the caveat that we have not shown that $c_0(n,l)$ are positive) that the
four point function with only the operators $(1,j_{\mu})$ in the S channel OPE, satisfies crossing symmetry
with unique values for the OPE coefficients and the anomalous dimensions (\ref{eq:allanomalous}).

\end{appendix}

\end{document}